\documentclass[10pt,conference,letterpaper]{IEEEtran}
\IEEEoverridecommandlockouts
\usepackage{cite}
\usepackage{amsmath,amssymb,amsfonts}
\usepackage{algorithmic}
\usepackage{graphicx}
\usepackage{textcomp}
\newcommand{\mycomment}[1]{}
\usepackage{xcolor}
    
\usepackage{physics}
\usepackage{epsfig}
\usepackage[capitalise]{cleveref}
\usepackage{amssymb}
\usepackage{svg}
\svgpath{{./figures/}}

\begin{document}

\title{LEO Clock Synchronization with Entangled Light\\
\thanks{Approved for Public Release: Distribution is Unlimited; $\#$23-0869.}
}

\author{
\IEEEauthorblockN{Ronakraj Gosalia\IEEEauthorrefmark{1}, Robert Malaney\IEEEauthorrefmark{1}, Ryan Aguinaldo\IEEEauthorrefmark{2}, Jonathan Green\IEEEauthorrefmark{2} and Peter Brereton\IEEEauthorrefmark{3}}
\IEEEauthorblockA{\IEEEauthorrefmark{1}University of New South Wales, Sydney, NSW 2052, Australia.}
\IEEEauthorblockA{\IEEEauthorrefmark{2} Northrop Grumman Corporation, San Diego, CA 92128, USA.}
\IEEEauthorblockA{\IEEEauthorrefmark{3} NASA Goddard Space Flight Center, Greenbelt, MD 20771, USA.}
}

\maketitle

\begin{abstract}
Precision navigation and timing, very-long-baseline interferometry, next-generation communication, sensing, and tests of fundamental physics all require a highly synchronized network of clocks. With the advance of highly-accurate optical atomic clocks, the precision requirements for synchronization are reaching the limits of classical physics (i.e. the standard quantum limit, SQL). Efficiently overcoming the SQL to reach the fundamental Heisenberg limit can be achieved via the use of squeezed or entangled light. Although approaches to the Heisenberg limit are well understood in theory, a practical implementation, such as in space-based platforms, requires that the advantage outweighs the added costs and complexity. Here, we focus on the question: can entanglement yield a quantum advantage in clock synchronization over lossy satellite-to-satellite channels? We answer in the affirmative, showing that the redundancy afforded by the two-mode nature of entanglement allows recoverability even over asymmetrically lossy channels. We further show this recoverability  is an improvement over  single-mode squeezing sensing, thereby illustrating a new complexity-performance trade-off for space-based sensing applications.
\end{abstract}


\section{Introduction}
In the pursuit of achieving a highly-synchronized network of clocks on low-Earth-orbit (LEO) satellites, a solution that is optimized for size, weight, and power (SWaP) is critical. Current state-of-the-art synchronization protocols, such as optical two-way time and frequency transfer \cite{Giorgetta2013}, suffer from a precision limit known as the standard quantum limit (SQL). Specifically, the scaling of temporal resolution with input resources is given by $\Delta t\propto 1/\sqrt{N}$ \cite{Giovannetti2001,Giovannetti2004}, where $N$ is the total number of photons exchanged during the measurement time. Fortunately, it is known that a better scaling is possible via the use of squeezed light or entangled light, both of which can theoretically reach a scaling of $\Delta t \propto 1/N$ \cite{Giovannetti2001,Giovannetti2004,Guo2020,Zhuang2020,Zhang2021}. This scaling is formally known as the Heisenberg limit. In this work, we refer to any timing measurement that, for a given $N$, is smaller than the SQL as a \emph{quantum advantage}.
On the technological roadmap, however, it is an open question as to whether any quantum advantage will be feasible in space under SWaP constraints.

In our recent work \cite{Gosalia2022}, we showed how a squeezed light setup on two LEO satellites could deliver a quantum advantage in clock synchronization despite typical sources of loss and noise such as beam diffraction, satellite-pointing jitter, and detector inefficiency. We conduct a similar assessment here, but this time with entangled light. Although, entanglement for clock synchronization has been extensively studied in the discrete variable regime before, in the form of single-photon entanglement (e.g. \cite{Jozsa2000,Okeke2018,Shi2022}), entanglement within the continuous variable regime has received relatively little attention. We investigate here the latter regime and study a system model to achieve quantum-enhanced clock synchronization between LEO satellites via the exchange of two-mode squeezed vacuum (TMSV) states. Through a detailed study that incorporates typical loss sources in LEO, we assess whether a quantum advantage exists in practice for such a system. We note the TMSV state is a workhorse for continuous-variable quantum communications \cite{Laudenbach2018,Hosseinidehaj2019,Villasenor2021}, interferometry \cite{Berchera2015}, and quantum computing \cite{Nielsen2010}. 

The remainder of this paper is as follows. In \cref{sec:methods} we detail our formalism, while \cref{sec:model} describes the system model. In {\cref{sec:sim}},  TMSV simulations are  discussed and in \cref{sec:related} a comparison with single-mode squeezed vacuum (SMSV) is provided. \cref{sec:conc} concludes our work. 

\section{Temporal modes}
\label{sec:methods}
The system model consists of a transmitter, Alice, who repeatedly emits picosecond-duration pulses in the TMSV state at a repetition frequency of order $100$~MHz synchronized to her clock. The receiver, Bob, mixes Alice's pulses with his coherent local oscillator (LO) pulses (having the same duration and repetition frequency as Alice) synchronized to his clock. Our objective is to synchronize Alice and Bob's clocks by measuring offsets between their pulses to a precision beyond the SQL. To achieve this objective, the temporal modes (TMs) of the TMSV state can be exploited --- a state originally centered around a single TM will excite higher order TMs when a non-zero offset exists \cite{Lamine2008}. In the following, we formalize the effect of offsets on a TM.

\subsection{Formalism}
In this section, we show that Alice's TMSV state can be fully described by a single TM. In our model, the TMSV state propagates in free space along the longitudinal $z$-axis with the following positive-frequency\footnote{Denoted in this work with a superscript $^{(+)}$.} 
electric field operator form \cite{Fabre2020}:
\begin{align}
    \label{eq:e+_orig}
    \hat{E}^{(+)}(u) = i\int_{-\infty}^{\infty}\sqrt{\frac{\hbar \omega}{2\epsilon_0}}\hat{a}(\omega)e^{-i\omega u} d\omega,
\end{align}
where $\hat{E}^{(-)}(u) = (\hat{E}^{(+)}(u))^\dag$ and $\hat{E}(u) = \hat{E}^{(+)}(u)+\hat{E}^{(-)}(u)$.
Here, $u= t-z/c$ is the spatial-temporal coordinate (also referred to as the mean light-cone variable \cite{Lamine2008}) along the $z$-axis and $t$-domain, $\omega$ is the angular frequency, $\hbar$ is the reduced Planck constant (we set $\hbar=2$ from \cref{sec:model} onwards), $\epsilon_0$ is the permittivity of free space, $c$ is the speed of light in vacuum, and $\hat{a}(\omega)$ is the monochromatic-mode annihilation operator. The monochromatic operators satisfy the commutation relation: $[\hat{a}(\omega), \hat{a}(\omega')] = \delta(\omega-\omega')$.
We can then define the $j$-th discrete annihilation superoperator $\hat{A}_j$ as $\hat{A}_j = \int_{-\infty}^{\infty}f_j^*(\omega) \hat{a}(\omega) d\omega$ where $\quad \hat{a}(\omega) = \sum_{j=0}^\infty f_j(\omega)\hat{A}_j$. Here, $f_j(\omega)$ is a weight function describing the spectral shape. Next, we apply a Fourier transform to obtain the time domain version: $\tilde{\hat{A}}_j = \int_{-\infty}^{\infty}\tilde{f}_j(u)\tilde{\hat{a}}(u)dt$, where $\tilde{f}_j(u) = \frac{1}{2\pi}\int_{-\infty}^{\infty}f_j(\omega)e^{-i\omega u}d\omega$ is the time-domain form of $f_j(\omega)$ and $\tilde{\hat{a}}(u) = (2\pi)^{-1}\int_{-\infty}^{\infty}\hat{a}(\omega)e^{-i\omega u}d\omega$.
Note, the weight functions are subject to the orthogonality relation \cite{Brecht2015}: $(2\pi)^{-1}\int_{-\infty}^{\infty}f_j^*(\omega)f_k(\omega)d\omega = \int_{-\infty}^{\infty}\tilde{f}_j^*(u)\tilde{f}_k(u)dt = \delta_{jk}$ and this imposes bosonic commutation on the superoperators such that $[\hat{A}_j, \hat{A}_k^\dag] = \delta_{jk}$. In this superoperator form, the electric field is a superposition over the entire TM set $\{v_j\}$:
\begin{align}
    \hat{E}^{(+)}(u) = \sum_{j=0}^\infty\hat{A}_jv_j(u),
    \label{eq:v_j}
    v_j(u) = i\int_{-\infty}^{\infty}\sqrt{\frac{\hbar\omega}{2\epsilon_0}}f_j(\omega)e^{-i\omega u}d\omega.
\end{align}
Notably, the current TM basis does not satisfy the orthogonality condition, that is $\int_{-\infty}^{\infty}v_j^*(u)v_k(u) du \neq 0, \forall j,k$ due to the factor of $\sqrt{\hbar \omega}$ in \cref{eq:v_j} \cite{Raymer2020}. However, making the assumption that the spectral profile is approximately monochromatic (i.e. $\omega \approx \omega_0$, where $\omega_0$ is the laser central frequency) provides a scenario where the orthogonality condition holds. Hence, $v_j(u) \approx \mathcal{E}\int_{-\infty}^{\infty}f_j(\omega)e^{-i\omega u}d\omega$, where $\mathcal{E} = i\sqrt{\hbar\omega_0/(2\epsilon_0)}$. Now, we define a new TM basis set, $\{y_\ell\}$, by applying a unitary transform: $y_\ell(u) = \sum_{j=0}^\infty \hat{U}_{\ell}^j v_j(u)$, with the fundamental mode \cite{Fabre2020,Treps2005} $y_0(u) = \sqrt{1/(\sum_{j=0}^\infty||v_j(u)||)}\sum_{j=0}^\infty v_j(u)$, where $||v_j(u)|| = \int_{-\infty}^{\infty}|v_j(u)|^2 du$. We denote the annihilation and creation operators for this new TM basis as $\hat{b}_\ell$ and $\hat{b}^\dag_\ell$, respectively, and modify \cref{eq:v_j} to:
\begin{align}
    \label{eq:e-plus}
    \hat{E}^{(+)}(u) = \mathcal{E}\hat{b}_0 y_0(u),
\end{align}
where the TMSV state is completely defined by the fundamental TM $y_0$.

\subsection{Spatial-temporal fluctuation}
Now, we want to find the effect an offset $\Delta u$ has on the TM formalism. Specifically, an offset will change \cref{eq:e-plus} to: $\hat{E}^{(+)}(u-\Delta u) = \mathcal{E}\hat{b}_0 y_0(u-\Delta u)$. Here, a Taylor series expansion about $u=0$ for small $\Delta u$ ($\Delta u \ll 1$~s) gives \cite{Lamine2008}:
\begin{align}
    \label{eq:e_deltau_final}
    \hat{E}^{(+)}(u-\Delta u) \approx \mathcal{E}\hat{b}_0\left(y_0(u) + \frac{\Delta u}{u_0}z_1(u)\right),
\end{align}
where $u_0 = \sqrt{1/(\omega_0^2+\Delta\omega^2)}$ is a normalization factor, $\Delta\omega$ is the frequency spread \cite{Lamine2008}, and
\begin{align}
    \label{eq:z_1}
    z_1(u) = \frac{1}{\sqrt{\Omega^2+1}}\left(y_1(u)+i\Omega y_0(u)\right),
\end{align}
denotes the timing mode, and $\Omega = \omega_0/\Delta\omega$. The expectation of \cref{eq:e_deltau_final} is \cite{Lamine2008}:
\begin{align}
    \langle\hat{E}^{(+)}(u-\Delta u)\rangle = \mathcal{E}\sqrt{N_{in}}e^{i\theta}\left(y_0(u)+\frac{\Delta u}{u_0}z_1(u)\right),
\end{align}
where $N_{in}$ is the total number of photons in Alice's state and $\theta$ is the global phase. Using the expectation we can equivalently express \cref{eq:e_deltau_final} as
\begin{align}
    \hat{E}^{(+)}(u-\Delta u) \equiv \mathcal{E}\left(\hat{b}_0y_0(u)+\hat{b}_1y_1(u)\right),
\end{align}
with $\langle\hat{b}_0\rangle=(1+i\omega_0\Delta u)\sqrt{N_{in}}e^{i\theta}$ and $\langle\hat{b}_1\rangle=(\Delta\omega\Delta u)\sqrt{N_{in}}e^{i\theta}$. 
Our analysis thus far shows that Alice's state can be fully described with TM $y_0$ when there is a zero offset ($\Delta u = 0$), and as a superposition of $y_0$ and the timing mode $z_1$ for small $\Delta u$. Bob can use the excitation of $z_1$ to measure $\Delta u$ by shaping his LO to $z_1$ \cite{Lamine2008}. We note here that an offset $\Delta u$ between Alice and Bob generalizes over the spatial domain $\Delta z$ and the temporal domain $\Delta t$, and for this reason we denote $\Delta u$ as the spatial-temporal offset.

\section{System model}
\label{sec:model}
\begin{figure*}
    \centering
    \includegraphics[width=\linewidth]{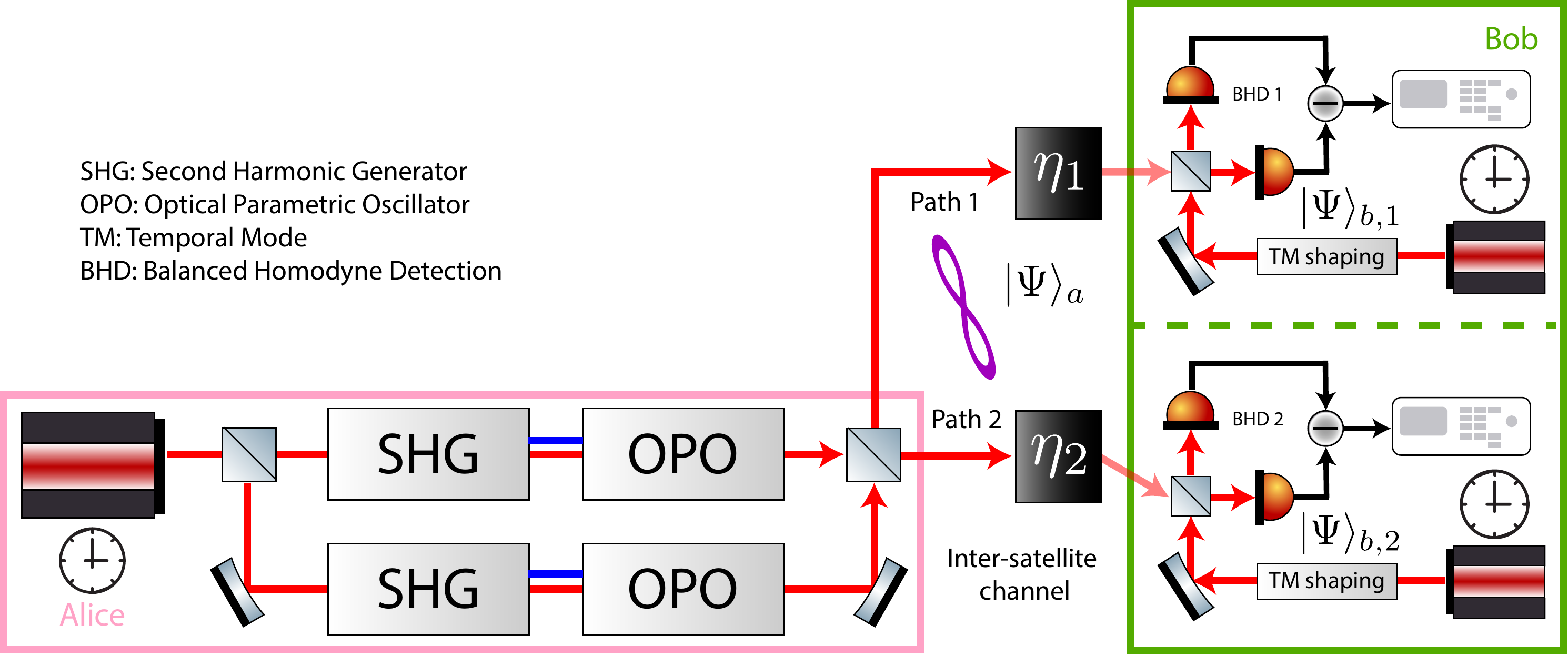}
    \caption{Alice, the transmitter in this one-way link, generates a TMSV state via a nonlinear down-conversion process consisting of an SHG and an OPO. The two modes of her TMSV state are emitted over spatially separate paths with effective transmissivities $\eta_1$ and $\eta_2$ to Bob. Bob mixes the TMSV state with his coherent state LO shaped in the TM $z_1$ and performs BHD. The BHD photocurrents are then post-processed to take advantage of the quadrature-correlations present in Alice's TMSV state and maximize the signal-to-noise ratio of the $\Delta u$ offset estimate. 
    }
    \label{fig:tmsv-setup}
\end{figure*}
The system model is shown in \cref{fig:tmsv-setup} where two one-way inter-satellite links between Alice and Bob are detailed. Alice emits the two modes of a TMSV state over separate spatial paths 1 and 2. At the receiver, Bob performs two sets of balanced homodyne detections (BHD) on the two modes of the TMSV state. Finally, on board post-processing is used to retrieve the variance of the correlated $\hat{X}$ quadrature thereby taking advantage of the entanglement within a TMSV state.
In the analysis to follow,  both BHD setups are assumed to be on the same satellite (Bob) with the measurement currents post-processed on-board. However, more generally, the two BHDs can be deployed on two separate satellites, each processing one of the two modes of Alice's TMSV state. The only additional assumptions we need for this case is that both receiver satellites are time synchronized, and can communicate their measurements over a classical link for post-processing. The analysis in this latter case would require only  slight modifications to that given here.

\subsection{Alice's entangled state}
At Alice, a second harmonic generator (SHG), an optical parametric oscillator (OPO), and a beam splitter (BS) are used to generate and coherently mix two single-mode squeezed vacuum (SMSV) states (squeezed in alternating quadratures $\hat{X}$ and $\hat{P}$), and generate a TMSV state. The SHG-OPO configuration is detailed further in \cite{Patera2010,Jiang2012,Lamine2008,Wang2018}. The TMSV state has correlated $\hat{X}$ quadratures and anticorrelated $\hat{P}$ quadratures that can be taken advantage of during measurement \cite{Lvovsky2015}. Although the setup used here is one of several ways to generate a TMSV state, the chosen approach allows for direct comparisons with an SMSV setup which is discussed later.

We simplify Alice's TMSV state representation by expressing it over only the first two TMs $\{y_0,y_1\}$ as follows:
\begin{align}
    \ket{\Psi}_{a} = \bigotimes_{\ell=0}^{1}\hat{D}_{\ell}(\sqrt{{N}_{1}}e^{i\theta_{1}})\hat{D}_{\ell}(\sqrt{{N}_{2}}e^{i\theta_{2}})\hat{S}_{\ell}(\xi)\ket{00}.
\end{align}
Here, $\hat{D}_{\ell}(\cdot)$ is the displacement operator in TM $\ell$, defined as $\hat{D}_{\ell}(\sqrt{{N}_{1}}e^{i\theta_{1}}) = \exp[\sqrt{{N}_{1}}(e^{i\theta_{1}}\hat{b}_{1,\ell}^\dag - e^{-i\theta_{1}}\hat{b}_{1,\ell})]$, where $\hat{b}_{1,\ell}$ is the annihilation operator in mode $1$ of the TMSV state and TM $\ell$, ${N}_1$ is the number of photons emitted from Alice to Bob along path 1, and $\theta_1$ is the corresponding phase. Similar expressions for path $2$ are found by substituting in index $2$. Further, $\hat{S}_{\ell}(\cdot)$ is the TMSV operator in TM $\ell$, defined as \cite{Lvovsky2015} $\hat{S}_{\ell}(\xi) = \exp\left(\xi\hat{b}_{1,\ell}^\dag\hat{b}_{2,\ell}^\dag-\xi^*\hat{b}_{1,\ell}\hat{b}_{2,\ell}\right)$, where $\xi = re^{i\phi}$ is the squeezing parameter and we set $\phi=0$ here to simplify $\xi=r$ as a real quantity. Finally, the operators are applied to the two-mode vacuum state $\ket{00}$.

\subsection{Bob and Charlie's balanced homodyne detection}
BHD is a technique used to optimally measure a specific quadrature of a signal by mixing it with a shaped LO. Here, in our system model, we mix each of the two modes of Alice's TMSV state separately with Bob's LO. We assume that the time of arrival of Alice's TMSV state is known \textit{a~priori} at Bob to an accuracy of order $10^{-12}$~s --- the technique we discuss here improves upon this accuracy.

Bob's LO are shaped in the TM $z_1$ and split into $\hat{E}_1$ and $\hat{E}_2$, where each have the positive-frequency electric field description:
\begin{align}
    \label{eq:e-Bob}
    \hat{E}^{(+)}_{1}(u) = \left(\hat{l}_0 y_0(u)+\hat{l}_1y_1(u)\right) = \hat{E}^{(+)}_{2}(u).
\end{align}
Here, we use the expanded form of $z_1$ from \cref{eq:z_1} and $\hat{l}_\ell$ denotes the LO annihilation operator in a TM $\ell$. For optimal detection, Bob's LO needs to be in a coherent state: $\ket{\Psi}_{b,1} = \bigotimes_{\ell=0}^{1}\hat{D}_{\ell}(\sqrt{{N}_{lo}}e^{i\theta_{lo}})\ket{0} = \ket{\Psi}_{b,2}$ (shown in \cref{fig:tmsv-setup}).
Here, we assume that BHD 1 and 2 both receive the same number of photons from the LO and have the same phase, and thus use the general variables ${N}_{lo}$ and $\theta_{lo}$. Similar to \cite{Lamine2008}, the expectation of \cref{eq:e-Bob} is:
\begin{align}
    \langle\hat{E}^{(+)}_{1}(u)\rangle = \frac{\mathcal{E}\sqrt{{N}_{lo}}e^{i\theta_{lo}}}{\sqrt{\Omega^2+1}}\left(i\Omega y_0(u)+y_1(u)\right) = \langle\hat{E}_2^{(+)}(u)\rangle,
\end{align}
where we have set $\langle\hat{l}_0\rangle = i\Omega\sqrt{{N}_{lo}/(\Omega^2+1)}e^{i\theta_{lo}}$ and $\langle\hat{l}_1\rangle = \sqrt{{N}_{lo}/(\Omega^2+1)}e^{i\theta_{lo}}$ \cite{Lamine2008}. 
As output, the BHD 1 measurement will generate a photocurrent given by $\hat{I}_{1} = |\mathcal{E}|^2\sum_{\ell=0}^1 \left(\hat{b}_{1,\ell}^\dag\hat{l}_{\ell} +\hat{b}_{1,\ell}\hat{l}^\dag_{\ell}\right)$ \cite{Lamine2008}. The expectation of the photocurrent is then
\begin{align}
    \nonumber \langle\hat{I}_{1}\rangle = 2|\mathcal{E}|^2\sqrt{{N}_{1} {N}_{lo}}\bigg[\frac{\Delta u}{u_0}\cos(\theta_1-\theta_{lo}) + \\
    \frac{\Omega}{\sqrt{\Omega^2+1}}\sin(\theta_1-\theta_{lo})\bigg],
\end{align} 
and a similar process would find that we can substitute in $N_2$ for $\langle\hat{I}_2\rangle$. We can also derive the variances  (equal for both photocurrents), which differ from \cite{Lamine2008} and can be written:
\begin{align}
    \langle\delta\hat{I}^2_{1}\rangle = |\mathcal{E}|^4 {N}_{lo}\cosh{2r} = \langle\delta\hat{I}^2_{2}\rangle.
\end{align} 
Although not required (as the individual timing measurements could be averaged directly), it will be useful to include a post-processing step whereby the individual photocurrents are added to give a final photocurrent: $\hat{I}_{\text{TMSV}} = \hat{I}_{1} + \hat{I}_{2}$. We will henceforth denote $\hat{I}_{\text{TMSV}}$ as the post-processed photocurrent. The \textit{ideal}\footnote{Here \textit{ideal} refers to the no loss scenario, and contrasts with \textit{real} where loss is considered.} post-processed photocurrent has expectation (assuming $\theta = \theta_{lo}$):
\begin{align}
    \label{eq:i-mean-ideal}
    \langle\hat{I}_{\text{TMSV},ideal}\rangle = \frac{2|\mathcal{E}|^2\Delta u\sqrt{{N}_{lo}}}{u_0}\left(\sqrt{{N}_{1}}+\sqrt{{N}_{2}}\right).
\end{align}
Similarly, with some algebra we can find the variance,
\begin{align}
    \label{eq:i-var-ideal}
    \nonumber \langle\delta\hat{I}_{\text{TMSV},ideal}^2\rangle = \langle\delta\hat{I}_{1}^2\rangle + \langle\delta\hat{I}_{2}^2\rangle + 2(\langle\hat{I}_{1}\hat{I}_{2}\rangle-\langle\hat{I}_{1}\rangle\langle\hat{I}_{2}\rangle) \\
    = 2|\mathcal{E}|^4{N}_{lo}\left(\cosh{2r}-\frac{\Omega^2-1}{\Omega^2+1}\cos(2\theta_{lo})\sinh{2r}\right).
\end{align}
Here, again assuming approximately monochromatic laser sources at Alice and Bob, $\Omega \gg 1$ (i.e. $\omega_0 \gg \Delta\omega$), and by setting the measurement phase to $\theta_{lo}=n\pi$ for $n\in\mathbb{Z}$ --- corresponding to an $\hat{X}$ quadrature measurement --- the final variance will be minimized to
\begin{align}
    \langle\delta\hat{I}_{\text{TMSV},ideal}^2\rangle = 2|\mathcal{E}|^4 {N}_{lo}e^{-2r}.
\end{align}
The variance is now phase-sensitive and reduces with increasing $r$ as desired. This reduced variance will minimize noise in our measurement of $\Delta u$ allowing us to reach a precision beyond the SQL. 

\subsection{Spatial-temporal offset estimate --- ideal}
Let us now derive $\Delta u$ in the ideal case by considering a signal-to-noise ratio (SNR) of 1 \cite{Lamine2008},
\footnote{In this work (similar to previous studies e.g. \cite{Lamine2008}) we set the signal-to-noise ratio to the worst-case of 1 --- the point where the signal is equal to the noise level, to find the minimum measurable spatial-temporal offset.}
where $\text{SNR} = \langle\hat{I}\rangle/\sqrt{\langle\delta\hat{I}^2\rangle}$ \cite{Lamine2008}. Using \cref{eq:i-mean-ideal} and \cref{eq:i-var-ideal}, we find
\begin{align}
    \Delta u_{\text{TMSV},ideal} = \frac{e^{-r}}{\sqrt{2}(\sqrt{{N}_1}+\sqrt{{N}_2})\sqrt{\omega_0^2+\Delta\omega^2}}.
\end{align}
Now, we assume photon conservation during the generation of the TMSV state and set $N_{in}=N_1+N_2$. Additionally, we assume that Alice's photons are evenly divided across the two modes of her TMSV state so $N_1=N_2$, giving
\begin{align}
    \label{eq:u-ideal}
    \Delta u_{\text{TMSV},ideal} = \frac{e^{-r}}{2\sqrt{{N}_{in}}\sqrt{\omega_0^2+\Delta\omega^2}}.
\end{align}
This $\Delta u_{\text{TMSV},ideal}$ represents the minimum offset that is measurable by a setup using TMSV state of $r$ squeezing level, $N_{in}$ photons, $\omega_0$ central frequency and $\Delta \omega$ frequency spread. However, so far we have not considered the effect of any losses and inefficiencies in the system.
\subsection{Spatial-temporal offset estimate --- realistic}
\label{sec:real}
We now extend our analysis to a realistic inter-satellite link by incorporating various sources of loss. We make use of the following BS unitary operator to model loss \cite{Lvovsky2015}
\begin{align}
    \label{eq:bs}
    \hat{\text{BS}} = \begin{bmatrix} \sqrt{\eta} & -\sqrt{1-\eta} \\ \sqrt{1-\eta} & \sqrt{\eta}\end{bmatrix},
\end{align}
where $\eta$ is the effective transmissivity, $0\leq\eta\leq1$, and loss is $1-\eta$. Our model encapsulates the effect of photonic loss due to beam diffraction, satellite pointing misalignment and detector inefficiencies into the general term $\eta$ (further details are in our previous work \cite{Gosalia2022}). The photocurrent in BHD 1 with BS operator applied is then
\begin{align}
    \nonumber \hat{I}_{1,real} = |\mathcal{E}|^2\sum_{\ell=0}^1\left(\sqrt{\eta_{1}}\hat{b}_{1,\ell}^\dag-\sqrt{1-\eta_{1}}\hat{v}_{1,\ell}^\dag\right)\hat{l}_\ell~+\\
    \left(\sqrt{\eta_{1}}\hat{b}_{1,\ell}-\sqrt{1-\eta_{1}}\hat{v}_{1,\ell}\right)\hat{l}_\ell^\dag.
\end{align}
Here, $\eta_1$ denotes the effective transmissivity along path 1, $\hat{v}$ [$\hat{v}^\dag$] denotes the annihilation [creation] operator of the vacuum state. $\hat{I}_{2,real}$ can be found by substituting in index $2$. The expectation of the \textit{real} post-processed photocurrent is then
\begin{align}
    \label{eq:i-mean-real}
    \langle\hat{I}_{\text{TMSV},real}\rangle = \langle\hat{I}_{\text{TMSV},ideal}\rangle\left(\sqrt{\eta_{1}}+\sqrt{\eta_{2}}\right)/2,
\end{align}
with variance given by
\begin{align}
    \label{eq:i-var-real}
    \nonumber \langle\delta\hat{I}_{\text{TMSV},real}^2\rangle = 2|\mathcal{E}|^4{N}_{lo}\bigg((\eta_{1}+\eta_{2})\sinh^2{r}+1+\\
    \sqrt{(1-\eta_{1})(1-\eta_{2})}-\sqrt{\eta_{1}\eta_{2}}\sinh{2r}\bigg).
\end{align}
Our analysis can be verified by setting $\eta_{1}=\eta_{2}=1$ in \cref{eq:i-mean-real} and \cref{eq:i-var-real} to retrieve \cref{eq:i-mean-ideal} and \cref{eq:i-var-ideal}, respectively. Finally, we derived $\Delta u$ with propagation loss at $\text{SNR}=1$:
\begin{align}
    \label{eq:u-real-tmsv}
    \Delta u_{\text{TMSV},real} =~&\frac{\sqrt{Q}}{  \left(\sqrt{\eta_{1}}+\sqrt{\eta_{2}}\right)\sqrt{{N}_{in}\left(\omega_0^2+\Delta\omega^2\right)}}, \\
    \nonumber Q =~&\left(\eta_{1}+\eta_{2}\right)\sinh^2{r}+1+\\
    \nonumber &\sqrt{\left(1-\eta_{1}\right)\left(1-\eta_{2}\right)}-\sqrt{\eta_{1}\eta_{2}}\sinh{2r}.
\end{align}
Again, setting $\eta_{1}=\eta_{2}=1$ retrieves the ideal case in \cref{eq:u-ideal}. Eq. (\ref{eq:u-real-tmsv}) is one of the main results of our study as it represents the minimum offset measurable by a setup using a TMSV state under lossy conditions.

In order to quantify the quantum advantage of our TMSV state, we note the SQL of a general single transmit/two receiver setup, which is found by setting $r=0$ in \cref{eq:u-real-tmsv}:
\begin{align}
    \label{eq:u-sql}
    \Delta u_{\text{SQL},real} = \frac{\sqrt{1+\sqrt{(1-\eta_1)(1-\eta_2)}}}{\left(\sqrt{\eta_{1}}+\sqrt{\eta_{2}}\right)\sqrt{{N}_{in}\left(\omega_0^2+\Delta\omega^2\right)}}.
\end{align}
Configurations that allow $\Delta u_{\text{TMSV},real} < \Delta u_{\text{SQL},real}$ have a quantum advantage and optimize the available $N_{in}$.
\section{Results}
\label{sec:sim}
\begin{figure}
    \centering
    \includegraphics[width=\columnwidth]{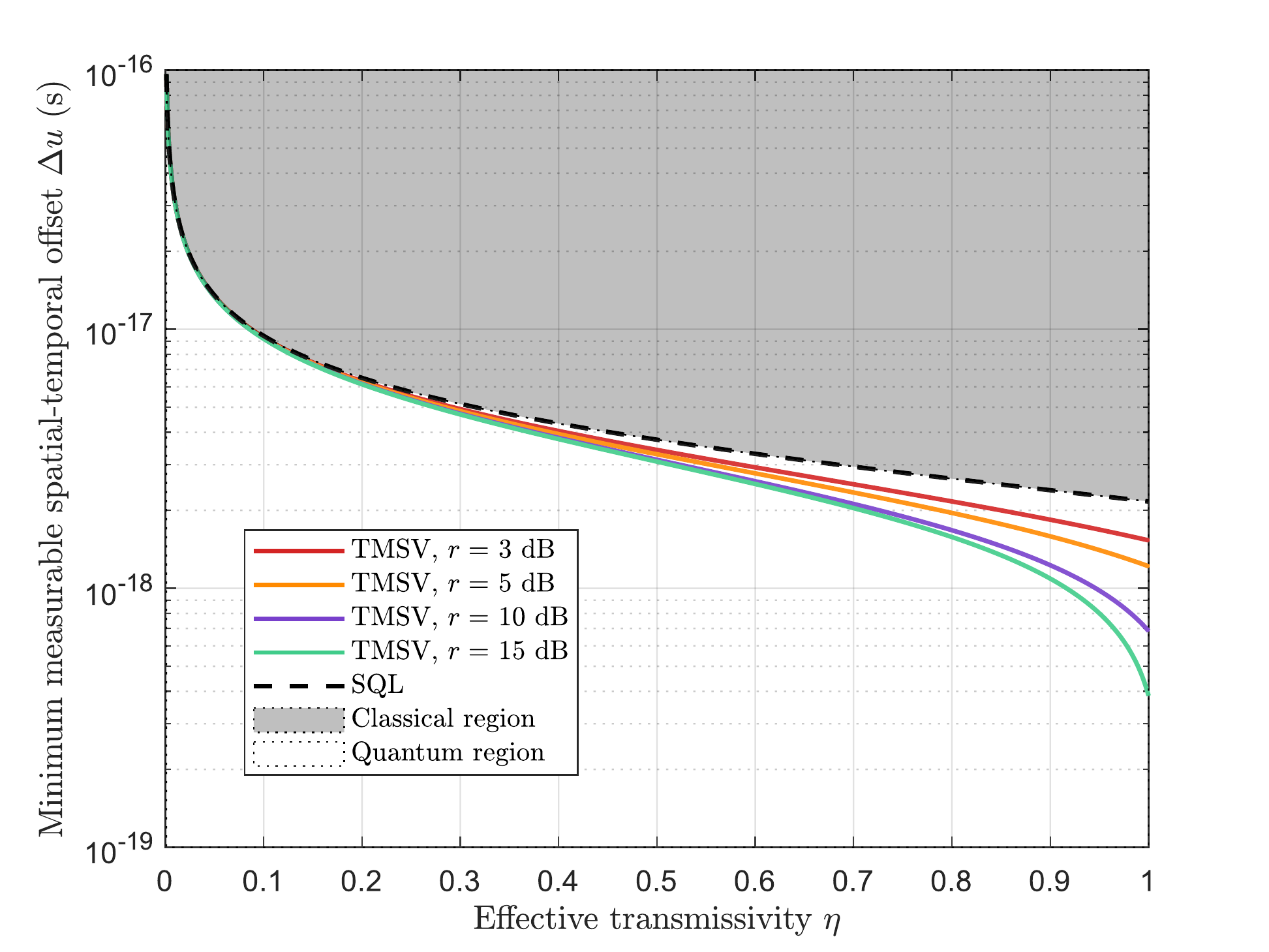}
    \caption{Minimum measurable spatial-temporal offset, $\Delta u$, vs. effective transmissivity, $\eta$. Here, $\eta_{1} = \eta_{2}\equiv\eta$, $\lambda_0=815$ nm, $\Delta\omega=2\pi\times10^6$ rad/s and $N_{in}=10^3$~photons. The gray highlights classical region with lower boundary being the SQL. Evidently, TMSV states with $3~$dB squeezing and more can achieve a quantum advantage since $\Delta u$ measurement below the SQL is possible when $\eta \gtrsim 0.4$. Notably, increasing the squeezing level is beneficial only when $\eta\gtrsim0.7$ to achieve even lower $\Delta u$.}
    \label{fig:tmsv-coh-comp}
\end{figure}
We now conduct a numerical simulation using system parameters representative of a LEO satellite link, namely: $\lambda_0=815$~nm ($\omega_0=2\pi c/\lambda_0$), $\Delta\omega=2\pi\times10^6$~rad/s, and $N_{in}=10^3$~photons. The chosen laser wavelength allows for use of SWaP-efficient silicon-based detectors \cite{Kingsbury2015}; while $N_{in}$ is 5 orders of magnitude smaller than the expected number of pump photons for a $10$~W $10$~ps laser system to represent generation inefficiency (discussed further in \cref{sec:related}). 

In the first simulation, Alice generates four different TMSV states with squeezing levels ranging from $r=3$~to $15$~dB\footnote{The squeezing parameter $r$ in dB is defined as $r_{\text{dB}} = -10\log_{10}\left(e^{-2r}\right)$.}. The measurable offset, $\Delta u$, is plotted for each $r$ and compared against the SQL in \cref{fig:tmsv-coh-comp}. Note, both path transmissivities are equal $\eta_1=\eta_2=\eta$. The results show a quantum advantage when $\eta \gtrsim 0.4$ since $\Delta u_{\text{TMSV},real}<\Delta u_{\text{SQL},real}$, even for small $r$. Additionally, increasing the squeezing level is advantageous only at high transmissivity ($\eta\gtrsim0.8$). Finally, in extremely lossy conditions, $\eta<0.3$, a quantum advantage does not exist as $\Delta u_{\text{TMSV},real}\approx\Delta u_{\text{SQL},real}$. 

\begin{figure}
    \centering
    \includegraphics[width=\columnwidth]{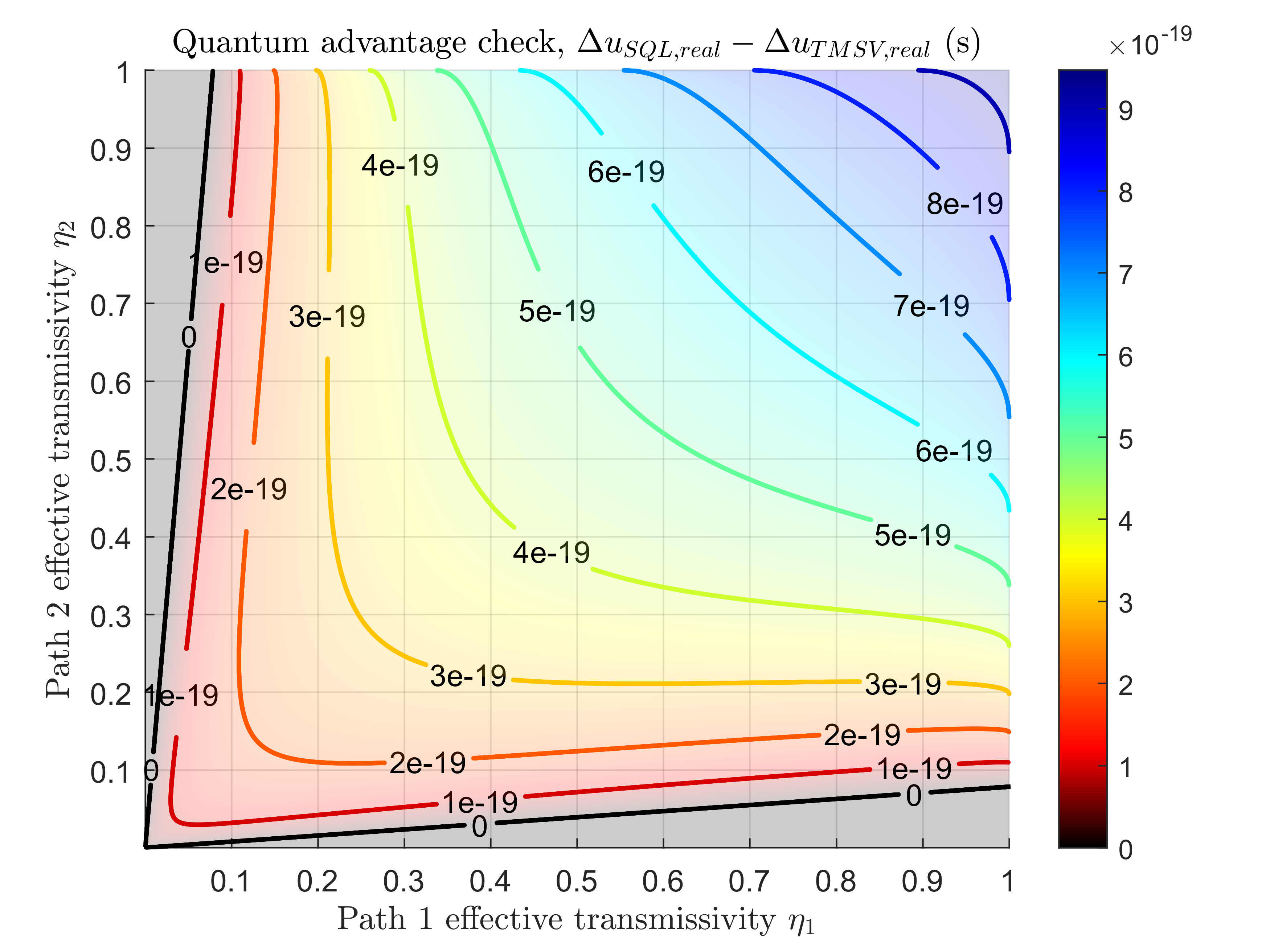}
    \caption{The quantum advantage check, $\Delta u_{\text{SQL},real} - \Delta u_{\text{TMSV},real}$, for various $\eta_1$ and $\eta_2$. Here, a constant squeezing level of $r=5$ dB is used and all other parameters are as per \cref{fig:tmsv-coh-comp}. Regions shown in color exhibit quantum advantage with contour lines indicating equal levels of quantum advantage. A minimum effective transmissivity of $\eta_{1},\eta_{2} \gtrsim 0.1$ is required.}
    \label{fig:tmsv-eta-comp}
\end{figure}
In the second simulation, we generalize the transmissivity terms considering $\eta_1$ and $\eta_2$ not equal (e.g two receiver satellites), and use a constant $r=5$~dB squeezing level. 
The results in \cref{fig:tmsv-eta-comp} show a quantum advantage exists as long as $\eta_{1},\eta_{2} \gtrsim 0.1$. The contour lines show regions with the same level of quantum advantage. For example, following the $6\times10^{-9}$~s contour line, when $\eta_1=\eta_2=0.695$ the quantum advantage is equal to when $\eta_1=0.585$ and $\eta_2=0.825$. Hence, a TMSV state can take advantage of asymmetrically lossy channels and compensate for one channel using another less lossy channel. 

Another example of the performance advantage over asymmetric channels is when $\eta_2=0.5$, $\Delta u_{\text{SQL},real}>\Delta u_{\text{TMSV},real}$ as long as $\eta_1>0.05$ --- this scenario shows that a quantum advantage is possible even in extremely lossy conditions. Although not visualized here, for $r>5$~dB, a larger quantum advantage is achievable, but at the cost of stricter requirements for $\eta_1$ and $\eta_2$. Overall, the unique advantage of the TMSV setup over asymmetrically lossy channels is not present in the SMSV setup in \cite{Lamine2008} where only a single channel exists, nor in a classical setup where the quadratures are uncorrelated. During LEO, channel transmissivity will be a dynamically varying parameter and thus this effect would be advantageous.

\section{Relation to other work}
\label{sec:related}
\begin{figure}
    \centering
    \includegraphics[width=\columnwidth]{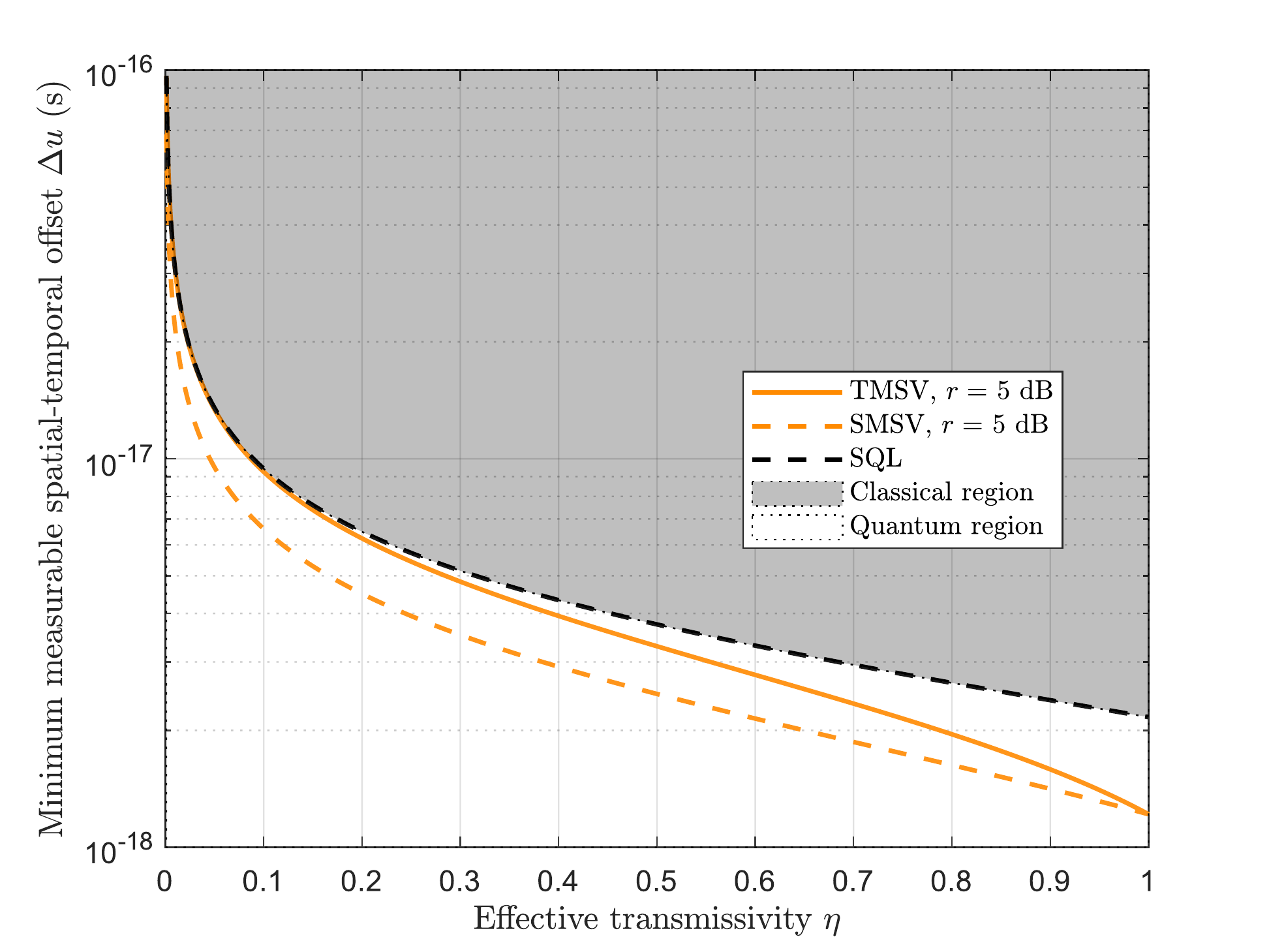}
    \caption{Minimum measurable spatial-temporal offset, $\Delta u$, vs. effective transmissivity, $\eta$, for setups with a TMSV and SMSV states. For equivalent (5~dB) squeezing, the SMSV state yields a lower $\Delta u$ than the TMSV state and thus a greater level of quantum advantage when $0.05<\eta<0.95$. Note, when $\eta=1$ both setups are equivalent and below the SQL.}
    \label{fig:tmsv-vs-smsv}
\end{figure}
We can now compare our analysis with previous studies in \cite{Lamine2008,Wang2018}, where an SMSV state over a one-way link was used instead. Our derivation in \cref{sec:model} reveals that $\Delta u_{\text{TMSV},ideal}\equiv\Delta u_{\text{SMSV},ideal}$, where the latter is from eq. (11) in \cite{Lamine2008}. This important finding shows that a TMSV state, in principle, yields the same level of precision (and in turn quantum advantage) as an SMSV state over an ideal channel. However, this equivalence in $\Delta u$ holds true only when all loss and noise effects are ignored. In contrast, we can find $\Delta u_{\text{SMSV},real}$ following the same process as before:
\begin{align}
    \label{eq:u-real-smsv}
    \Delta u_{\text{SMSV},real} = \frac{1}{2}\sqrt{\frac{\eta_1 e^{-2r}+(1-\eta_1)}{\eta_1{N}_{in}(\omega_0^2+\Delta\omega^2)}}.
\end{align}
A third simulation is then conducted with results summarized in \cref{fig:tmsv-vs-smsv}. For $r=5$~dB squeezing, we find $\Delta u_{\text{SMSV},real} < \Delta u_{\text{TMSV},real}$ when $0.05\lesssim\eta\lesssim0.95$. Hence, SMSV in fact outperforms TMSV over symmetrically lossy channels.

Although a TMSV state exhibits advantageous performance over asymmetrically lossy channels, however, the trade-offs in SWaP between both states further complicate the matter. In our analysis, equal power was maintained by using equal $N_{in}$ across the approaches considered, since laser energy is proportional to the number of photons emitted. Additionally, we assume that the efficiency of converting $N_{in}$ to $N_{1}$ and $N_{2}$ is equal for the SMSV and TMSV states since the same SHG-OPO optical system is used. However, the TMSV state requires one additional SHG-OPO path for the second mode which adds to the required SWaP. Additionally, a co-located second receiver aperture and post-processing stage are needed for the TMSV state. Hence, the additional SWaP requirement for a TMSV state needs to be traded-off with any expected gains over asymmetric channels in practice.

Furthermore, a classical setup may have higher pulse transmit efficiency than both the SMSV and TMSV states due to a simpler optical system. For example, current 3U CubeSats have a $10$~W optical power budget \cite{Kingsbury2015}, which over a $10$~ps duration pulse with $\lambda_0=815$~nm is approximately $4.1\times10^8$~pump photons/pulse. In a TMSV state, these pump photons are spontaneously down-converted via the SHG-OPO setup into signal and idler photons in a probabilistic manner. In our study, we integrated the effect of this probabilistic conversion process by setting $N_{in} = 10^3$~photons. However, in a classical state, where there is no parametric down-conversion, the pump photons convert directly into signal/idler photons with much higher efficiency, and thus have a higher effective transmit power. Having said this, our analysis does ensure a fair comparison by using the same $N_{in}$ for the SQL, TMSV and SMSV approaches.

\section{Conclusion}
\label{sec:conc}
In this study, we presented an analysis of a space-based quantum-enhanced synchronization system using entangled light. We derived the minimum measurable spatial-temporal offset, $\Delta u$, for a TMSV state over lossless and lossy channels. Also, by comparing with the SQL, we found practical scenarios where the quantum approach is indeed advantageous. The simulation results presented show system and channel configurations that enable a quantum advantage at squeezing levels as low as $3$~dB for constant $N_{in}$. Our results also found a unique advantage of a setup using a TMSV state whereby redundancy in the two-modes allows for recoverability over asymmetrically lossy channels. This redundancy supersedes both the classical and SMSV approaches and would be useful on inter-satellite links. Finally, a comparison of sensitivity between the SMSV and TMSV states reveals equivalence in performance under ideal channel conditions, as well as the superior performance of SMSV in certain inter-satellite channels. 
This last point highlights a new trade-off
with regard to performance versus implementation complexity. 



\bibliographystyle{IEEEtran}
\bibliography{references}

\end{document}